\documentclass[aps,prb,epsf]{revtex4}
\usepackage{graphicx}
\usepackage{epsfig}

\def\c82{{C$_{82}$}}
\def\gdc82{{Gd@\c82}}
\newcommand{\ba}{\begin{eqnarray}}
\newcommand{\ea}{\end{eqnarray}}
\newcommand{\be}{\begin{equation}}
\newcommand{\ee}{\end{equation}}
\newcommand{\bea}{\begin{eqnarray}}
\newcommand{\eea}{\end{eqnarray}}
\def\etal{{\it et al}.}

\def\eqnref#1{Eq.(\ref{#1})}
\def\sectref#1{Section \ref{#1}}

\begin{document}

\def\CC{{\rm\kern.24em \vrule width.04em height1.46ex depth-.07ex \kern-.30em C}}

\title{Electronic transport, structure, and energetics of endohedral \gdc82  metallofullerenes}

\author{L. Senapati, J. Schrier, and  K. B. Whaley}

\affiliation{ Department of Chemistry and Pitzer Center for Theoretical Chemistry,\\University of California, Berkeley, CA 94720-1460, USA.
}

\begin{abstract}
  
Electronic structure and transport properties of the fullerene \c82
and the metallofullerene \gdc82 are investigated with density
functional theory and the Landauer-Buttiker formalism. 
The ground state structure of \gdc82 is found to have the
Gd atom below the C-C bond on the C$_2$ molecular axis of \c82.
Insertion of Gd into \c82 deforms the carbon chain in the vicinity of
the Gd atoms. Significant overlap of the electron distribution is
found between Gd and the \c82 cage, with the transferred Gd electron
density localized mainly on the nearest carbon atoms. This charge
localization reduces some of the conducting channels for the
transport, causing a reduction in the conductivity of the \gdc82
species relative to the empty \c82 molecule.  The electron transport
across the metallofullerene is found to be insensitive to the spin
state of the Gd atom.

\end{abstract}

\maketitle

\section{Introduction}

Endohedral metallofullerences \cite{Shinohara00} have attracted wide
interest due to their functional characteristics and potential
applications in the field of nanomaterials and biomedical
science.\cite{BJS+93} Recently, metallofullerene doped nanotubes
("peapods") have also attracted experimental attention due to their
structural and electronics properties,\cite{LKK+02,HSB+00,SOT+02} and
have been proposed as a possible self-assembled quantum computing
architecture.\cite{AAB+03} Additionally, recent experimental studies
of \gdc82 in single wall carbon nanotube ((\gdc82)@SWNT) peapods have
shown novel transport behavior.\cite{OSS+03} The first step in
understanding the properties of the peapod structures is an adequate
treatment of the individual \gdc82 fullerenes.

In 
earlier \gdc82 theoretical calculations by Kobayashi and Nagase,\cite{KN98} the
relaxation of the fullerene cage was not considered, nor
the possibility of novel geometries for the Gd atom in the fullerene
cage was not considered; the predicted ground state spin multiplicity
M = 9, resulting from ferromagnetic coupling between the Gd f-electrons and the
odd electron on the fullerene cage, is in contrast to
electron spin resonance 
determinations of a M=7 ground state,
due to antiferromagnetic coupling of $-1.8\,{\rm cm}^{-1}$.\cite{FOK+03}
The M=7 ground state is also supported by magnetic
studies.\cite{FSO+95,FSYT95, HYZ99,HYZ00}
 In order to resolve this discrepancy,
 we present results of density functional calculations for
the equilibrium geometries, magnetic and transport properties of
\gdc82, made with the Landauer-Buttiker transport formalism.  
\sectref{methods} gives a brief description of the theoretical procedure, \sectref{results} present our results, and \sectref{conclusion} draws conclusions and discusses
possible future directions.
   
\section {Computational Methods}\label{methods}

The equilibrium geometry and the total energy of \gdc82 are calculated
using density functional theory (DFT).\cite{ParrYang} One of the
primary considerations involved in these calculations is determination
of a suitable basis set and exchange correlation
functional. We therefore compared the bond length, binding
energy and spin multiplicity values of Gd and Gd dimer (Gd$_2$)
optimized using several different basis sets and exchange correlation
functionals, to experimental values\cite{KL71} and theoretical
all-electron relativistic calculations\cite{DLK00}.  As we discuss in
\sectref{atomic}, the results are quite sensitive to the
choices of both basis set and exchange correlation functional. A
relativistic effective core potential CEP-121G \cite{CS93} for the
atoms with the generalized gradient approximation of Becke's exchange
functional\cite{Becke93} and Lee-Yang-Parr correlational
functional\cite{LYP88} (B3LYP) gave results comparable to experiment,
and were thus used for the subsequent calculations.

The total energies and forces are calculated using the linear
combination of atomic orbitals (LCAO) molecular orbital
approach.\cite{ParrYang} We employed a 6-31G basis for the carbon
atoms and a CEP-121G basis set for Gd. The computations were performed
using Gaussian03 for various spin multiplicities.\cite{Gaussian03}
The geometries were optimized without symmetry constraints by
minimizing the total energy and requiring the forces to vanish within a
10$^{-3}$
a.u./Bohr threshold, 
 at
every atom site.
 To find the ground state structure for \gdc82, initial
positions of the Gd atom inside the \c82 cage were generated, then
optimized without constraining any of the atomic coordinates. The
calculations were repeated for different spin multiplicities to
determine the ground state spin configuration, as we discuss in \sectref{structure}.  Although the partitioning of charge to specific atoms or groups within
a molecule is not uniquely defined within the postulates of quantum
mechanics,\cite{SO96} a recent experimental evaluation of density
functional charge schemes\cite{YSB04} found the commonly used Mulliken
charge analysis\cite{SO96} to be generally deficient for charge
calculations, and Natural Population Analysis\cite{RWW85} (NPA) to be
the most generally accurate method of those studied.

To study the transport, we created a model comprised of \c82 located
between two Au contacts, with and without the endohedral Gd atom. The
numbers of Au atoms on each side of the contact were varied to study
the transport dependence on contact geometry. Fig.1 shows one of these
structures with a triangular contact of 3 Au layers in which the layer
adjacent to \c82 has one Au atom, the second has 3 Au atoms and the
third layer has 6 Au atoms.  For calculation of electron transport
through \gdc82 and \c82, we used the non-equilibrium Green's function
based Landauer-Buttiker formalism.\cite{Datta,TDH+98,PSAN03} Neglecting
spin-flip processes, the total current due to coherent scattering is
given by $I_{spin-coherent}$=$I{^\alpha}$ + $I{^\beta}$ , where
${I^\alpha}$ and ${I^\beta}$ are the contribution to current from spin
up ($\alpha$) and down ($\beta$) states, respectively.  The spin
specific contributions are expressed as the integral over the
injection energy of the tunnelling electron, $E$,
\be \label{lb-eqn}
I^{\alpha(\beta)} = (e/h){\int}^{\mu_1}_{\mu_2}
T^{\alpha(\beta)}(E,V)[f(E,\mu_1) - f(E,\mu_2)]dE ,
\ee
taken between the electrochemical potentials $\mu_1$ and $\mu_2$,
\bea\label{mu}
\mu_1 &=& E_f - ({q_1} V/(q_1+q_2))  \\
\mu_2 &=& E_f + ({q_2} V /(q_1+q_2)),
\eea
where $q_1$ and $q_2$ are charge accumulated in
the left and right contacts respective, $V$ is the applied bias, and  
$E_f$  is the Fermi energy of
the gold contact, taken as -5.53 eV for our calculation. 
 ${f(E,\mu)}$ is the
Fermi distribution function and ${T^{\alpha(\beta)}(E,V)}$ is the
transmission function that represents the sum of transmission
probabilities for electrons of a given spin through the
contact/metallofullerene complex, obtained from the overlap and
Hamiltonian matrices determined by the DFT calculation.  The explicity
included Au atoms in our model system described above are used to
obtain the coupling matrices for calculation of self-energy
functions.\cite{Datta,TDH+98} We have used the local density of states
of the 6s-band of bulk gold (0.035 eV per electron spin) to
approximate the Green's function of the (bulk) Au
contact.\cite{TDH+98}

\section{Results} \label{results}
\subsection{ Properties of Gd and Gd$_2$ }\label{atomic}

First, we discuss the geometry, ionization potential, and electron
affinity of Gd and Gd$_2$ and compare these with experiment. This
comparison is provided to assess the accuracy of our theoretical
procedure.

To assess how well our method accounts for the properties of the Gd
atom, we have calculated the ionization potential of Gd and Gd$_2$ as
well as the binding energy and Gd$_2$ dimer bond length. The ground
state Gd atom was found to have a spin multiplicity of 9 for all
functionals that we have considered with the CEP-121G effective core
potential basis functions.\cite{CS93} Results for the
LSDA,\cite{ParrYang} B3LYP\cite{Becke93,LYP88} and PW91\cite{PW92}
functionals are presented in Supplemental Table I.  The calculated
binding energy value (1.7595 eV) and bond-length (2.926 \AA) for
Gd$_2$ with CEP-121G as frozen core basis function and B3LYP as
functional for spin multiplicity of 19 agrees well with the
experimental binding energy (1.784$\pm$0.35 eV)\cite{KL71} and the
relativistic all-electron theoretical bond-length (2.895
\AA)\cite{DLK00}.  The ionization potential for Gd and Gd$_2$ were
calculated as 5.014 eV and 4.119 eV respectively.  Based on these
results, the CEP-121G basis set for Gd and 6-31G basis set for C
atoms, together with the B3LYP functional, were used for the
Gd@C$_{82}$ calculations.

Although we have employed a frozen core, the Gd 5s and 5p core
electrons are allowed to relax in the molecular calculations. Atomic
calculations show that the relaxation of these states is often crucial
for an accurate determination of valence state properties because of
the overlap of the 5d electron orbitals with the other n=5 orbitals.

\subsection{ Structure of \gdc82}\label{structure}

We have taken a \c82 fullerene cage with C$_{2v}$ point group symmetry
 \c82(C$_{2v}$) and optimized the structure with Gd inside at various
 positions, as described in \sectref{methods}. In the \c82(C$_{2v}$)
 fullerene cage, the C$_2$-axis goes through the center of one
 six-membered ring and one C-C double bond. Our calculations reveal
 that the ground state structure of \gdc82 has the Gd atom situated
 adjacent to the C-C double bond on the symmetry axis, as shown in
 Fig. 2. This is in contrast to reported results for other metal atoms
 such as Sc, Y and La in
 \c82,\cite{TUN+95,KN98,NTS+00,NTS+98,KN02,KN03} where the metal atoms
 are found to be centered inside the fullerene cage on top of the
 C$_{2v}$ axis six-membered ring.\cite{KN98,NTS+00} Recent synchrotron
 radiation powder structure analysis experiments of \gdc82 are
 consistent with our calculated placement of the Gd atom.\cite{NIS+04}
 The distances between Gd and the two nearest C atoms are 2.38 \AA\,
 and 2.41 \AA\, respectively. The binding energy of Gd to \c82 is
 found to be 5.6435 eV (130.1 kcal/mole), comparable to those for the
 Sc, Y, and La metallofullerenes;\cite{KN98} the energy levels of \c82
 and \gdc82 are shown in Supplemental Figure 1. Addition of Gd into
 \c82 slightly lengthens the proximal C-C bond (from 1.425 \AA\, to
 1.473 \AA) atom, as well as the neighboring C-C bonds (by 0.01 to
 0.04 \AA).  This deformation results in an energy difference between
 the \c82(C$_{2V}$) symmetry structure and the {\gdc82} optimized
 deformed structure when recalculated without Gd of 0.109 eV (2.51
 kcal/mole).

As discussed in \sectref{methods}, the NPA charge partitioning tends
to give better results; our \gdc82 calculations agree with this
analysis, as the Mulliken charge analysis indicates charge transfer of
1.43 electrons from Gd atom to the C$_{82}$ cage, and NPA indicates a
charge transfer of 2.43 electrons.  The latter value agrees well with
electron energy loss spectroscopy (EELS) experiments.\cite{SIKS00} The
electrons are localized near the two proximal carbon atoms, resulting
in strong charge density overlap between the Gd atom and these C
atoms.  NPA analysis also gives a Gd $d$-orbital population of 0.48
electrons, consistent with a Dewar-Chatt back-bonding interaction
between the Gd atom and the carbon double bond.\cite{Crabtree} The
deformation of the C-C bonds near the Gd is also consistent with
back-bonding, as the back donation of charge from the C-C bonds to the
d-orbitals weakens (and hence lengthens) the bonds.  Additionally, the
Gd $f$-orbital population of 7.02 is consistent with the charge
transfer from the cage of 0.04 electrons empirically determined by
Nada\"i \etal\, for reproducing X-ray magnetic circular dichroism
experiments.\cite{NMD+04}

In order to study the difference in Gd placement, as compared to the
Sc, Y, and La cases, we performed single point calculations, in which the
Gd atom was placed in the La-like position, adjusting the distance
between the hexagon and the Gd atom to account for the smaller radius
of the Gd${}^{+3}$ ion as compared to the La${}^{+3}$
ion;\cite{CottonWilkinson} The lowest total energy structure we found
occured when the Gd atom was placed 2.10 \AA\, above the plane of the
hexagon; this structure had a total energy of 0.104 hartrees (2.83 eV)
higher than the ground position determined by our optimizations.  NPA
analysis determined nearly identical charge distribution on the Gd
atom in both cases, attesting to the importance of Gd-C back-bonding
in stabilizing the structure.  However, an elementary electrostatic
analysis, using the NPA atomic charges as point ions,
gives a Gd-cage potential energy of -1.015 and -0.9366 hartrees
for the ground and La-like positions respectively; the difference
between these nearly accounts for the difference in total energies. 
This
interpretation is consistent with resonant photoemission
spectroscopy experiments which suggest that the Gd-C bond is primarily
ionic in character.\cite{PSC+04}

We have also examined the effect of different spin configurations on
the optimized the \gdc82 structure. The ground state structure was
found to have 7 unpaired electrons on Gd and one unpaired electron on the
\c82 cage aligned antiparallel to the Gd spins, with total spin
multiplicity M=7, in agreement with magnetic studies.\cite{FSO+95,FSYT95, HYZ99,HYZ00}
 The next lowest energy structure has M=9, with seven
unpaired electrons at Gd and the odd \c82 cage electron oriented
parallel to the Gd spins. The energy difference between these two
configurations is 2.6 meV ($16.1\,{\rm cm^{-1}}$), which compares
favorably to the experimentally determined value of $14.4\, {\rm
cm^{-1}}$.\cite{FOK+03} The detailed energies with various spin
multiplicities are listed in Table I.  


\subsection{ Electronic transport}\label{transport}

In order to see the effects of contact geometry, we have taken between
  one and three layers of Au atoms to represent the contact.  To
  obtain the contacts, single Au atoms were placed on each side of the
  \gdc82 molecule, shown schematically in Fig. 1, at a distance of
  2.015 \AA\, from the carbon atoms, along the C$_{2v}$ symmetry axis.
  To build the larger contacts, the bond lengths and geometry of bulk
  Au were used to add 3 Au atoms to the second layer and 6 Au atoms to
  the third layer, as shown in Fig. 1.  Although it is known that the
  conductance can be strongly dependent on the contact
  structure,\cite{KBY04} this transport model serves as the basis for
  understanding the qualitative transport properties of \gdc82.

Calculated I-V characteristics for \c82 and \gdc82 with various spin
states and Au contact geometries are shown in Figs. 3, 4, 5,
and Supplemental Fig. 3. The ground state (M=7) transport with a
single Au atom contact on each side of the fullerene is shown in
Fig. 3. Addition of the Gd atom into the \c82 cage leads to a
reduction in the current. To understand this difference, we computed
the density of state (DOS) for both structures, and found the
metallofullerene to have a reduced DOS near the Fermi energy as
compared to \c82.  These results are shown in Supplemental Fig. 2.
The I-V characteristics for the two- and three-layer contact structures were 
found to be 
  qualitatively the same as for the single Au atom contact
  case, and are shown in Supplemental Fig. 3 and Fig. 4, respectively. 
Quantitatively, the current is
seen to increase with the addition of further gold
layers, due to the increasing number of conduction channels.

Calculations for the 
  M=9 structure, which is structurally degenerate to the M=7 system, are shown
  in Fig. 5. We find only a small difference in current between the
  two spin configurations, and the conduction properties of
  spin up and spin down electrons (as calculated in \eqnref{lb-eqn}) 
were found to be insensitive to the 
Gd spin configurations, suggesting that the spin of the
  Gd is shielded by the cage carbon atoms and that transport occurs
  consequently only through the fullerene cage. This is also supported
  by plots of the highest occcupied and lowest unoccupied molecular
  orbitals (HOMO/LUMO), shown in Figs. 6 and 7, which indicate that conduction
  occurs primarily through the \c82 cage and shows no signature of conducting
  paths involving the Gd atom. 
This suggests that
incorporation of Gd decreases the number of available conduction
channels between the contacts, as a result of the charge localization
effect noted in \sectref{structure}.

By attaching
the Au contacts, one expects some charge transfer from the metallic
contact to the fullerene at zero bias, but this effect (which is the
same for both structures) is relatively small due to the weak coupling
between the fullerene and metal contact.  
Using the NPA method described in the previous section, the charge
transfer from the left and right Au contacts to the \c82
cage were found to be 0.09 and 0.11 electrons for single layer of
Au, 0.12 and 0.13 electrons for the two layer contact and 0.13
and 0.15 electrons for the three layer contact.  
Use of these charge values in the expression for the electrochemical
potential, \eqnref{mu}, had no effect on the calculated
transport properties.  Despite the fact that Gd is closer to one of the
electrodes, we observed no diode like behavior; we attribute this to
the the conduction occuring primarily through the fullerene cage, which is 
approximately symmetrical with respect to both contacts.

\section{Conclusions}\label{conclusion}

We have analyzed the energetics, structure, and transport properties
of \gdc82. The present results indicate that the CEP-121G basis set
for Gd, with the B3LYP density functional for exchange and
correlation, is a satisfactory combination for DFT calculations for
large Gd endofullerene systems. Our calculated structure for \gdc82
confirms the geometry determined by x-ray diffraction data, indicating
that the Gd is located adjacent to the C-C double bond on the symmetry
axis, in contrast to the other Group-3 metallofullerenes.
Furthermore this
geometry gives calculated spin configurations that agree with ESR
and magnetic experiments, in contrast to previous theoretical studies.  
Using the Landauer-Buttiker formalism for
transport, we find that conduction occurs primarily through the \c82
cage, and that charge donation from the Gd atom to the cage disrupts
these conduction channels. We find no evidence for electron-spin
dependent transport effects due to the spin state of the Gd
atom.  Besides providing a solid methodological and structural basis
for future calculations of {(\gdc82)@SWNT} structures, the absence of
HOMO/LUMO density on the Gd atom---and thus relative lack of
interaction with the surrounding nanotube---helps explain why the $M_4$
and $M_5$ peak edges of the {(\gdc82)@SWNT} EELS spectrum were found
to be identical to that for \gdc82, indicating that the Gd valence state 
is unaffected by the surrounding nanotube.\cite{OSS+03}

\section{Acknowledgements} 

J.S. thanks the National Defense Science and
Engineering Grant (NDSEG) program and U.S. Army Research Office
Contract/Grant No. FDDAAD19-01-1-0612 for financial support.
K.B.W. thanks the Miller Institute for Basic Research in Science for
financial support.  This work was also supported by the Defense
Advanced Research Projects Agency (DARPA) and the Office of Naval
Research under Grant No. FDN00014-01-1-0826, and the National Science
Foundation under Grant EIA-020-1-0826. We thank the National
Computational Science Alliance for partial support under the grant
No. DMR030047.


\vfill\eject

\begin{table}
\caption{Calculated total energies and binding energies for the optimized structures of \c82 and \gdc82  as a function of Gd spin multiplicity, M.  The M=7 state of \gdc82 is found to be 2.6 meV below the M=9 state in total energy.}
\begin{tabular}{|c|c|c|} \hline\hline
${Atom(M)}$& ${TE(au)}$& ${BE (eV)}$  \\ 
\hline
\c82(C$_{2v}$)      &-3123.805927 & \\
\gdc82(M=5)        &-3235.8783289  &4.4066  \\
\gdc82(M=7)        &-3235.9237819  &5.6435 \\
\gdc82(M=9)        &-3235.9236904  &5.6409 \\
\gdc82(M=11)       &-3235.87890696 &4.4224 \\
\hline\hline
\end{tabular}
\end{table}

\begin{figure}[ht]
\epsfig{figure=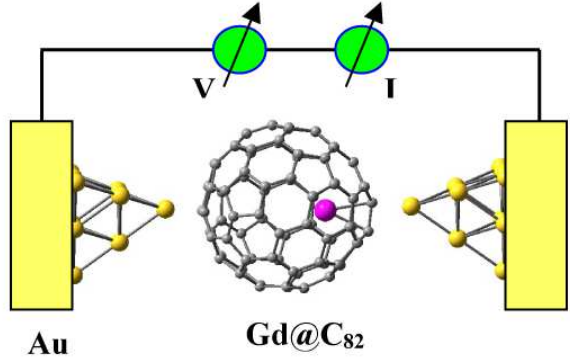,width=8.5cm}
\caption{Gold contact model used for transport calculations across
\gdc82 (see part b). The first layer adjacent to \c82 has one Au atom,
the second has 3 Au atoms and the third layer has 6 Au atoms.  }
\end{figure}

\begin{figure}[ht]
\epsfig{figure=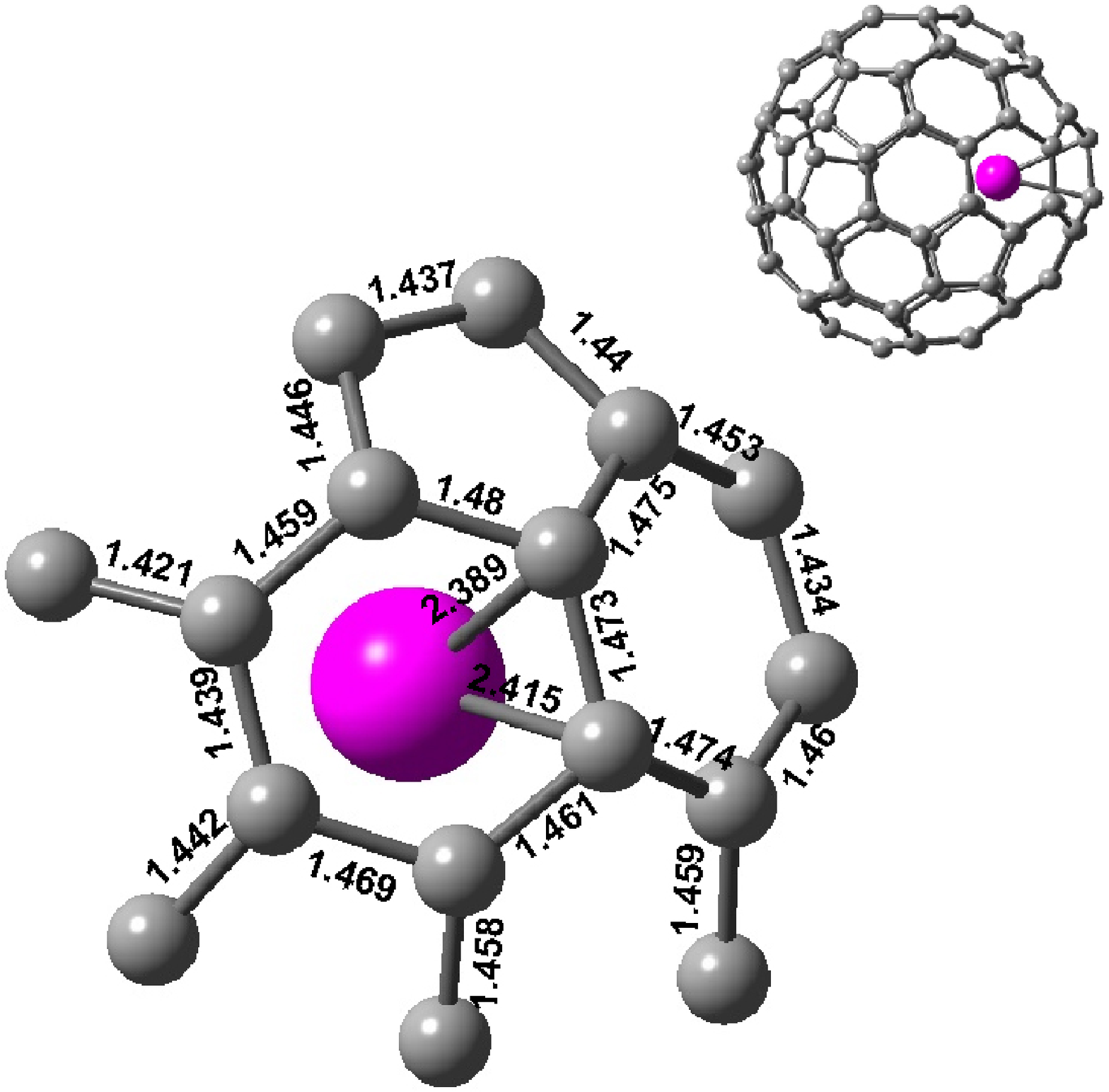,width=8.5cm}
\vspace{0.5cm}
\caption{Optimized structure of \gdc82. The ground state structure of
\gdc82 has the Gd atom (purple) situated adjacent to the C-C double
bond on the C2v symmetry axis that connects a six-membered ring with a
C=C bond on the other side of the fullerene. The distances between Gd
and the two nearest C atoms are 2.38 \AA\, and 2.41 \AA\,
respectively. The lines between Gd and C are used to indicate the
closest carbon atoms.  }
\end{figure}

\begin{figure}[ht]
\epsfig{figure=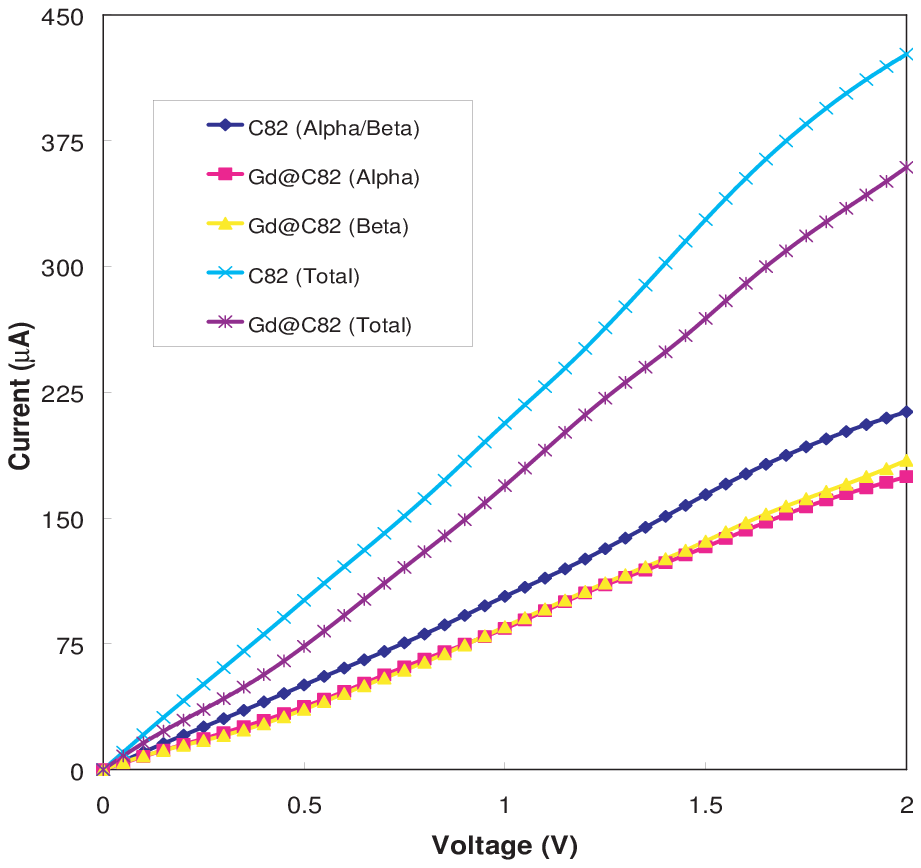,width=8.5cm}
\caption{I-V plot for single Au atom contacts.  Total conduction (sum
of spin up and spin down conduction) for \gdc82 is lower at all bias
voltages than for pure fullerene \c82. Conduction of spin up (alpha)
and spin down (beta) electrons is similar at all bias values.  }
\end{figure}

\begin{figure}[ht]
\epsfig{figure=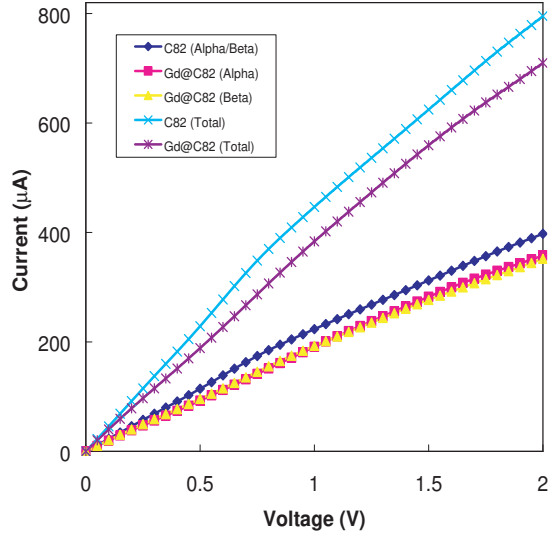,width=8.5cm}
\caption{I-V plot for contact geometry shown in Figure 1.  Note the
qualitative similarity to the other contact geometries (Figs. 3,
Supplement Fig. 2).  }
\end{figure}

\begin{figure}[ht]
\epsfig{figure=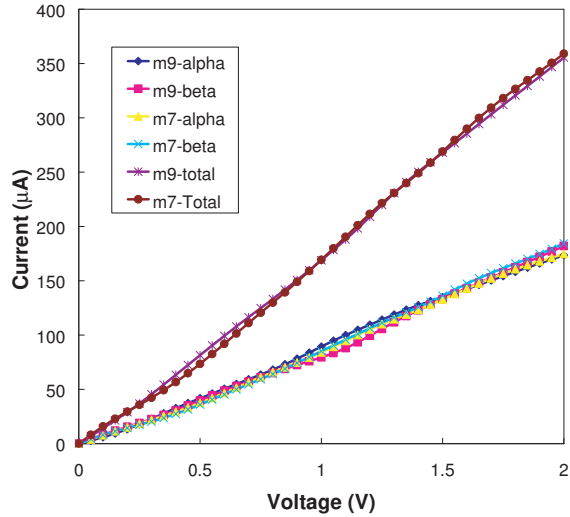,width=8.5cm}
\caption{I-V plot for conduction in the spin multiplicity M=7 ground
state, and the M=9 statee with energy 2.6 meV higher in energy.
Conduction characteristics for spin polarized and unpolarized currents
are similar for both cases, with neither showing strong spin-dependent
transport effects.  This suggests that the current is mainly carried
through the \c82 cage, and does not involve the Gd atom.  }
\end{figure}

\begin{figure}[ht]
\epsfig{figure=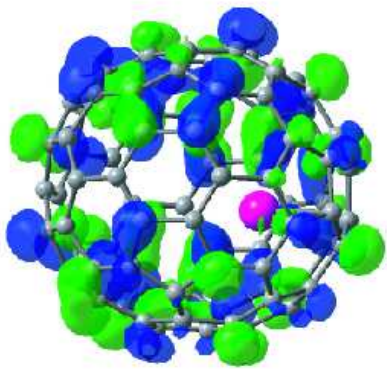,width=8.5cm}
\caption{Lowest Unoccupied Molecular Orbital (LUMO) plot of \gdc82 showing localization on the \c82  cage without significant LUMO orbitals on  the Gd atom. Blue and green are used to indicate the positive and negative sign of the wavefunction, respectively.  The purple sphere indicated the Gd atom.
}
\end{figure}

\begin{figure}[ht]
\epsfig{figure=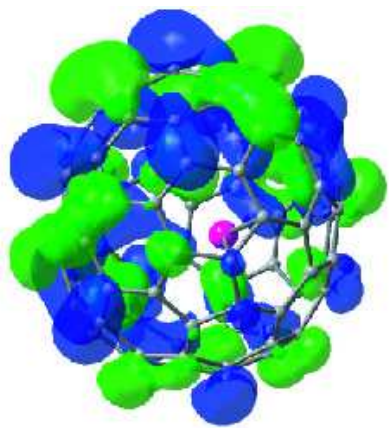,width=8.5cm}
\caption{Highest Occupied Molecular Orbital (HOMO) plot of \gdc82.  Blue and green are used to indicate the positive and negative sign of the wavefunction.  HOMO orbitals are localized on cage of \c82 and there is no HOMO orbital on Gd (purple sphere). From both HOMO and LUMO plots, it is evident that the current is mainly carried through the \c82 cage, as opposed to the Gd atom.
}
\end{figure}

\end{document}